# What drives passion?
# An empirical examination on the impact of personality trait interactions and job environments on work passion

Annika Breu[1,2], Taha Yasseri[2,3,4,5]*

[1] TUM School of Management, Technische Universität München, Munich, Germany
[2] Oxford Internet Institute, University of Oxford, Oxford, UK
[3] School of Sociology, University College Dublin, Dublin, Ireland
[4] Geary Institute for Public Policy, University College Dublin, Dublin, Ireland
[5] Alan Turing Institute for Data Science and AI, London, UK

*Corresponding Author:
Taha Yasseri, School of Sociology, University College Dublin, Belfield, Dublin D04 V1W8, Republic of Ireland. *taha.yasseri@ucd.ie*

**Abstract**
Passionate employees are essential for organisational success as they foster higher performance and exhibit lower turnover or absenteeism. While a large body of research has investigated the consequences of passion, we know only little about its antecedents. Integrating trait interaction theory with trait activation theory, this paper examines how personality traits, i.e. conscientiousness, agreeableness, and neuroticism impact passion at work across different job situations. Passion has been conceptualized as a two-dimensional construct, consisting of harmonious work passion (HWP) and obsessive work passion (OWP). Our study is based on a sample of N = 824 participants from the myPersonality project. We find a positive relationship between neuroticism and OWP in enterprising environments. Further, we find a three-way interaction between conscientiousness, agreeableness, and enterprising environment in predicting OWP. Our findings imply that the impact of personality configurations on different forms of passion is contingent on the job environment. Moreover, in line with self-regulation theory, the results reveal agreeableness as a "cool influencer" and neuroticism as a "hot influencer" of the relationship between conscientiousness and work passion. We derive practical implications for organisations on how to foster work passion, particularly HWP, in organisations.

*Keywords*: Big Five, work passion, trait interaction, trait activation, self-regulation

## INTRODUCTION

Worldwide, only 13% of workers are passionate about their work (Gallup, 2013). Passion at work embodies "a strong inclination toward an activity that people like, that they find important, and in which they invest time and energy" (Vallerand & Houlfort, 2003, p. 177). This emotional investment helps organisations to thrive (Gallup, 2013). In the rapidly changing global economy, work passion becomes increasingly decisive for organisations that aim to sustainably increase labour productivity. However, the current literature has largely neglected the antecedents of work passion rather focusing on the consequences of passion at work such as performance (Astakhova, 2015; Astakhova & Porter, 2015; Burke, Astakhova, & Hang, 2015; Ho, Wong, & Lee, 2011), work satisfaction and turnover intentions (Houlfort, Philippe, Vallerand, & Ménard, 2014), or product innovation (Klaukien, Shepherd, & Patzelt, 2013). More recent work has shown evidence for social contagion of work passion (Ho, Garg, & Rogelberg, 2021).

In the extant literature, two types of passion are usually distinguished: Harmonious work passion (HWP) and obsessive work passion (OWP) (Vallerand & Houlfort, 2003). The two passion types differ regarding how they absorb passionate work. Employees with HWP keep in control of their work and have a high work satisfaction (Burke et al., 2015) whereas workers with OWP easier lose control over their work, and are at higher risk for burnout (Vallerand, Paquet, Philippe, & Charest, 2010). In addition to this, there are more recent efforts to redefine and operationalize work passion (Chen, Lee, & Lim, 2020).

Extant theory on work passion has suggested that personality may be a major antecedent of work passion (Vallerand & Houlfort, 2003). Thus, we aim at advancing our understanding of how personality drives passion. A few existing studies on this topic have started to explore the main effects of personality traits on passion (Balon, Lecoq, & Rimé, 2013; Tosun & Lajunen, 2009). However, emerging research in trait interaction theory suggests that testing different trait constellations (e.g., interaction of conscientiousness and extraversion) may yield a more comprehensive picture about the influence of personality on work-related outcomes ( Judge & Erez, 2007; Witt, Burke, Barrick, & Mount, 2002). More recent work based on electronic performance monitoring at a small scale reported the positive impact of consciousness and extraversion on work passion (Hussain et al., 2021).

Whiles the role of job type and personality traits have been well studied, what remains as a gap is the interactions between personality traits in the contexts of different work environments. To address the current gap in the literature on how personality traits interact in predicting passion at work, our paper focuses on trait interactions. As self-control and passion are closely interlinked, we use self-regulation theory to predict the direction of the trait interaction (Metcalfe & Mischel, 1999; Ode, Robinson, & Wilkowski, 2008). Accounting for situational specificity improves the validity of single traits (Tett & Jackson, 1991; Barrick & Mount, 1991) as situational cues activate specific traits (Tett & Burnett, 2003; Tett & Guterman, 2000). To consider the impact of job environments on trait interactions to predict work passion, the RIASEC taxonomy classifies different job environments (Holland, 1959; 1985). Our study aims at fostering our understanding of what drives passion at work by considering the impact of trait interactions in different job environments on work passion. We make three major contributions to the literature.

First, we advance the passion literature by investigating how individual differences influence work passion (Perrewé, Hochwarter, Ferris, McAllister, & Harris, 2014). Focusing on trait interactions extends previous passion research and allows for a finer-grained insight into the impact of different trait constellations on passion at work.

Second, we contribute to trait activation theory by testing the activating effect of job environments on trait interactions to predict work passion (Tett & Burnett, 2003). By identifying jobs that activate trait interactions we advance the current state since prior research has mostly focused on single traits and their interaction with the environment (Barrick & Mount, 1991; Tett, Jackson & Rothstein, 1991). However, prior research suggests that trait interactions are important for understanding how configurations of personality traits interact with the environment in predicting work outcomes (Barrick & Mount, 2005; Penney et al., 2011).

Third, we build on previous trait interaction research (Ode et al., 2008) and apply self-regulation theory (Metcalfe & Mischel, 1999) to predict the direction of trait interactions. Further, our study demonstrates the potential of using online data sources with social science research. Since the data is drawn from the myPersonality project (Kosinski & Stillwell, 2016), a Facebook application with a focus on psychometric personality tests among a range of other questionnaires, we could take advantage of a large-scale data source drawn from a sample diverse in education, race, job, country, or political opinion.

**Passion at work**

Building on self-determination theory (Deci & Ryan, 1985; 2000), Vallerand et al. (2003) developed the concept of dualistic passion with the two dimensions of harmonious passion and obsessive passion. One of the motivational processes defining self-determination is the process of internalisation that adapts behaviours to one's identity (Ryan & Deci, 2000). This dualistic conceptualization also applies to passion at work (Vallerand and Houlfort, 2003).

Individuals who are harmoniously passionate about their work experience an autonomous internalisation that creates a strong sense of volition (Deci & Ryan, 2000). They value their work as important as they voluntarily chose this job. Driven by their true interest, they gain pleasure from work. As they are in control of their job, their work does not profoundly interfere with other activities in their life enabling a successful work-life balance (Vallerand et al., 2003). When unable to do their work, employees with HWP can relax from work. During work, however, HWP enables a high concentration on the task, flow, and work satisfaction (Vallerand et al., 2010).

In contrast, OWP results from a controlled internalisation and is fuelled by intra- or interpersonal pressures (Deci & Ryan, 2000). Individuals who are obsessively passionate about their work consider their job to be highly central in their lives, for instance, due to social or organisational acceptance, salary increase, or promotions (Astakhova, 2015). Since such internal forces take over control of the person`s behaviour, employees are unable to completely disengage from thinking about work, which prevents full focus and inhibits work satisfaction (Mageau & Vallerand, 2007; Ratelle et al., 2004; Vallerand & Houlfort, 2003).

Consequently, HWP relates to a more positive work experience displayed in higher psychological well-being (Houlfort et al., 2014), facilitated psychological adjustment (Bélanger et al., 2015), and decreased burnout risk (Vallerand et al., 2010) compared to OWP. While OWP negatively relates to psychological work outcomes, its association to work performance is similarly positive as for HWP (Burke et al., 2015; Ho et al., 2011).

Overall, our article investigates the relationship between personality configurations (i.e., trait interactions) and the job context on the one hand, and HWP and OWP on the other hand. In the following, we review extant evidence and derive our hypotheses. An overview of our theoretical model and hypothesis is displayed in Figure 1.

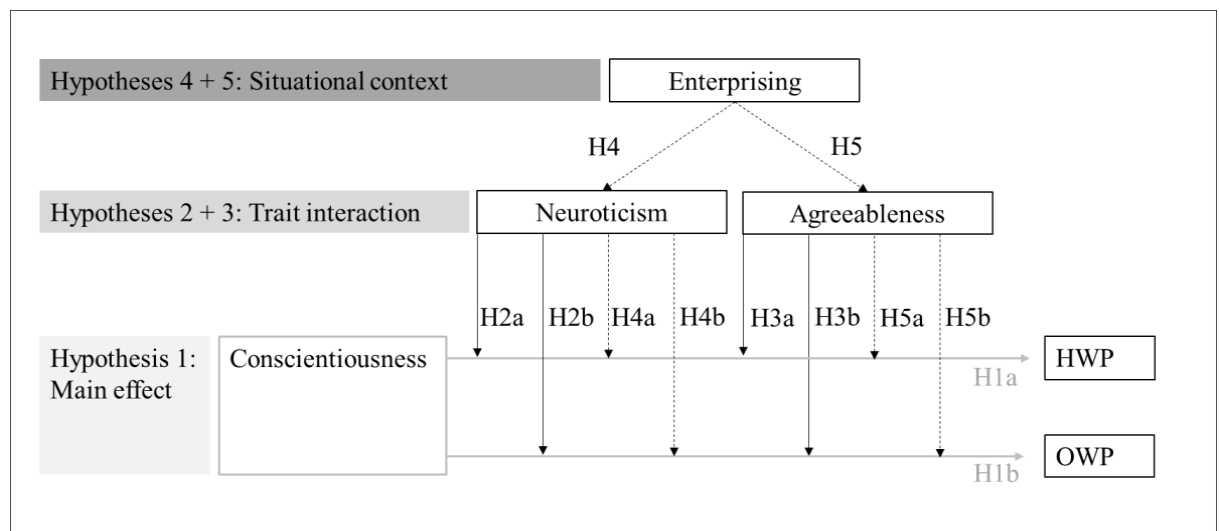

Figure 1: Illustration of the expected interactions of main effects, trait interactions and job environments.

**The relationship between the Big Five personality traits and passion**

To understand the underlying motivational forces of passion, we focus on the role of personality and personality configurations (Vallerand et al., 2003; Vallerand & Houlfort, 2003). The five-factor model (FFM) captures the complex construct of personality with five personality factors: Extraversion, agreeableness, conscientiousness, neuroticism, and openness to experience (e.g. Fiske, 1949; Goldberg, 1990; 1992; Digman, 1990). Goldberg framed the label "Big Five" for these five factors (1981, p. 159). Personality research that determines the impact of personality on work passion is very scant. The few studies focusing on personality to predict passion either used other personality frameworks than the FFM, such as the Eysenck's Personality Questionnaire-Revised (EPQ-R), to operationalise personality (Tosun & Lajunen, 2009) or targeted not specifically work passion (Balon et al., 2013). In response to this state of the art, scholars have called for research that advances our understanding of how personality affects passion (Perrewé et al., 2014).

HWP and OWP differ in their relation to self-control, one superordinate dimension of the Big Five traits (Olson, 2005). Prior research derived superordinate dimensions by factoring out two superordinate dimensions – self-control and engagement – from the FFM (Carroll, 2002; Olson, 2005; Digman, 1997). Harmoniously passionate individuals can control their work passion, whereas obsessively passionate individuals find it difficult and often lose control of their work (Vallerand & Houlfort, 2003). This implies that self-control characteristics are decisive in differentiating between HWP and OWP and may contribute to explain the internalisation of work passion. Self-control strongly correlates with conscientiousness, agreeableness, and neuroticism. Conscientiousness has shown the strongest associations with constraint (McCrae & Löckenhoff, 2010) and its relation to work passion is in the focus first.

Conscientious employees comply with "socially prescribed impulse control" that supports delaying gratification, thinking before acting, and following rules (John & Srivastava, 1999, p. 121). Two facets of conscientiousness related to constraint capture these characteristics: Some name them premeditation and perseverance (Whiteside & Lynam, 2001), others label them deliberation and self-discipline (Costa & McCrae, 1992). The first facet entails the ability to not concede to first impulses and think before actions follow. The second facet refers to the ability to force oneself to reach aspired goals (McCrae & Löckenhoff, 2010).

Regarding work passion, the high self-control abilities may enable conscientious employees to keep control over their work passion such that work stays in harmony with other activities in life. This fits the definition of HWP in that employees excel at work but can release thoughts after work (Vallerand & Houlfort, 2003). With the ability to control the passion for work, conscientiousness may be in favour of HWP. The self-disciplined and deliberative behaviour of conscientiousness contrasts OWP which implies difficulties to control the job (Vallerand & Houlfort, 2003). Even though conscientious employees show strong persistence in even tedious tasks, deliberative forces may prevent OWP. Thus, we hypothesise:

    **H1a:** Conscientiousness is positively related to HWP.
    **H1b:** Conscientiousness is negatively related to OWP.

**Big Five trait interactions and passion**

The few existing studies examining the impact of personality traits on passion have overwhelmingly concentrated on the direct effects of personality traits (Balon et al., 2013; Tosun & Lajunen, 2009). However, prior research indicates that trait interactions exhibit enhanced predictive power compared to individual traits in studies examining work outcomes in the organizational context such as performance (e.g. Guay et al., 2013; Warr, Bartram, & Martin, 2005), counterproductive work behaviour (CWB) (e.g. Jensen & Patel, 2011; Zhou et al., 2014), volunteering (Carlo, Okun, Knight, & Guzman, 2005), helping behaviour (King, George, & Hebl, 2005), and occupational stress (Grant & Langan-Fox, 2006). Self-regulation theory helps to determine the interaction of conscientiousness with agreeableness and neuroticism to predict work passion.

In his classical work, Freud (1927) described self-regulation as reflecting on the constant conflict between the *id* and the *ego*, i.e., between the temptation to give in impulses and a rational force who inhibits doing so. Metcalfe & Mischel (1999) systematically studied these two processes and coined them as "hot systems" and "cool systems". Hot systems respond to external stimuli in an emotional, impulsive, and reflexive way. Cool systems are characterized by cognitive, flexible, and integrated responses. The hot *go* system impedes self-control whereas the cool *know* system is the base for self-control (Metcalfe & Mischel, 1999). Metcalfe & Mischel (1999) proposed that both hot and cold processes exist in the brain and interact with each other.

Ode et al. (2008) predicted the direction of interactions between agreeableness and neuroticism on anger and aggression using the hot-system/cool-system framework. They proposed neuroticism to be a *hot* or impulsive influence, and agreeableness to be a *cool* or inhibiting influence concerning anger and aggression (Ode et al., 2008). We build on these findings of neuroticism as a hot influencer, and agreeableness as a cool influencer. We extend previous research by explicitly considering the interaction between agreeableness and neuroticism and how these traits influence conscientiousness to predict work passion.

**Neuroticism as a hot influencer.** Neurotic individuals often have nervous, sad, and tense feelings (John & Srivastava, 1999) accompanied by low self-esteem and rigid perfectionism (McCrae & Costa, 1999). Moreover, neuroticism exhibits dysfunctional self-control that manifests in either under-control (Tangney, Baumeister, & Boone, 2004; Whiteside & Lynam, 2001) or over-control (Schnabel, Asendorpf, & Ostendorf, 2002). Under-control implies a feeling of urgency (Whiteside &

Lynam, 2001) that results in impulsive behaviour. Over-control originates from anxiety or high self-consciousness that leads to inhibited behaviour and potentially to compulsive actions (McCrae & Löckenhoff, 2010). This reveals that self-control is more difficult for neurotic individuals (McCrae & Löckenhoff, 2010).

Studies exploring the interaction between conscientiousness and neuroticism found that given high conscientiousness, emotionally stable employees were more likely to engage in helping behaviour than highly neurotic employees (King et al., 2005). Moreover, the interaction was significant in predicting CWB on the organisational or individual level (Jensen & Patel, 2011; Bowling, Burns, Stewart, & Gruys, 2011): Least CWB occurred among highly conscientious and low neurotic employees. This pattern remained significant in a triple interaction with organisational constraints (Zhou et al., 2014).

The dysfunctional self-control associated with neuroticism may prevent HWP that demands an intact self-control system (Vallerand & Houlfort, 2003). While conscientiousness may foster HWP due to its self-discipline and deliberation skills (e.g. McCrae & Löckenhoff, 2010), the dysfunctional self-control abilities of neuroticism (e.g. Whiteside & Lynam, 2001) may reduce HWP. On the other hand, OWP is reactive to impulses and associated with low self-control capabilities (Vallerand & Houlfort, 2003). The dysfunctional self-control of neuroticism may foster OWP whereas conscientiousness may decrease OWP such that neuroticism may undermine the self-control capabilities of conscientiousness. We hypothesise:

**H2a:** The positive relationship between conscientiousness and HWP will be stronger for low neuroticism than high neuroticism.

**H2b:** The negative relationship between conscientiousness and OWP will be stronger for low neuroticism than high neuroticism.

**Agreeableness as a cool influencer.** Agreeableness also showed associations to self-control (Tangney et al., 2004) or behavioural inhibition (Smits & Boeck, 2006) in prior research. Agreeable individuals are at ease absorbing desirable social rules, which results in courteous, thoughtful behaviour and the ability to control aggression. On the contrary, low agreeableness makes controlled behaviour less likely and neglecting social rules fosters antagonistic behaviour (McCrae & Löckenhoff, 2010). Regarding the interaction between conscientiousness and agreeableness on performance, supervisor ratings of workers high in both traits exceeded those of highly conscientious but low agreeable workers (Guay et al., 2013). This pattern was held for helping behaviour, where highly conscientious highly agreeable individuals were more likely to help colleagues than their less agreeable co-workers (King et al., 2005). Further, only a high level of both traits reduced CWB (Jensen & Patel, 2011).

The strong but not excessive self-control capabilities of agreeableness (e.g. Tangney et al., 2004) may strengthen HWP. Thus, both conscientiousness and agreeableness may positively influence HWP because of their strong relation to self-control skills. This implies that agreeableness may strengthen the hypothesised positive relation between conscientiousness and HWP. However, the ability of agreeableness to control emotions and not give in temptations immediately (Tangney et al., 2004) contrasts the low self-control associated with OWP (Vallerand & Houlfort, 2003) and may reduce the risk of OWP. Due to their strong self-control abilities, both agreeableness and conscientiousness may weaken OWP; we thus hypothesise:

**H3a:** The positive relationship between conscientiousness and HWP will be stronger for high agreeableness than low agreeableness.

**H3b:** The negative relationship between conscientiousness and OWP will be stronger for high agreeableness than low agreeableness.

**How trait interactions influence passion across job environments**

**Trait activation theory.** Meta-analytic evidence suggests higher predictive validity of trait interactions when models of personality trait interactions include the job situations they are contextualized in (Zhou, Meier, & Spector, 2014). Responding to outstanding calls for research to include situational moderators (Barrick & Mount, 2005; Penney et al., 2011; Tett & Christiansen, 2007) to advance our understanding of how job environments impact how trait interactions influence passion, we include job environments as situational moderators.

Trait activation theory considers how situations interact with personality traits (Tett & Burnett, 2003). The main idea of trait activation is an interactionist paradigm by positing that the expression of traits responds to specific trait-relevant situational cues (Tett & Guterman, 2000). Tett & Burnett (2003) proposed three trait-relevant situational cues at the task, social, and organisational levels. For this study, we focus on the task level which reflects individuals' job demands. We thereby respond to research calls to explore the impact of task-level cues on personality (Penney et al., 2011). Also, investigating job demands on the task level is important due to their predictive relevance in the context of employee selection (Tett & Burnett, 2003).

**Holland's RIASEC model.** The RIASEC taxonomy is a common framework operationalising job demands on the task level (Tett & Burnett, 2003). Holland clustered occupational interests based on job descriptions in the Dictionary of Occupational Titles (DOT) and found six major work environments: realistic (R), investigative (I), artistic (A), social (S), enterprising (E), and conventional (C), in short RIASEC. The result of Holland's classification is a code with the first letter of three environments whereby the letter order determines which abilities and interests are most important for this job (Holland, 1959; 1985). *Consistency* captures the overlap or internal coherence between RIASEC environments and *differentiation* refers to the extent that two RIASEC environments are distinct from each other (Holland, 1985). Based on studies that explore relations between the Big Five and preference for RIASEC jobs (Fruyt & Mervielde, 1999), Tett & Burnett (2003) showed which RIASEC environments provide cues for the activation of traits: conventional, enterprising, artistic and investigative environments activate conscientiousness, whereby enterprising, social and realistic environments activate the trait agreeableness, and neurotic traits are activated in enterprising, conventional, realistic and investigative environments.

To increase the predictive validity of the trait interactions in predicting work passion, RIASEC environments ideally activate all three traits that are in the focus of this paper. Only the enterprising environment activates all three traits conscientiousness, agreeableness, and neuroticism (Fruyt & Mervielde, 1999; Tett & Burnett, 2003). Therefore, the enterprising environment is in the focus in the following. The enterprising environment has been described as an environment where both social and technical skills are required and where success and reputation are admired (Holland, 1985). People successful in this environment demonstrate a high level of social as well as technical skills filled with ambition and desire for success.

However, while the activating power of the enterprising environment holds for the individual traits, the effect on trait interactions is unclear. To derive hypotheses, we checked if also trait interactions become active in Enterprising environments. After classifying samples of previously explored trait interactions into RIASEC environments, we focused on interactions between conscientiousness and agreeableness (C*A) or neuroticism (C*N).

**Trait interactions in the Enterprising environment.** Trait interactions occurred primarily in enterprising environments (e.g. Guay et al., 2013; Warr et al., 2005). We observed the impact of the enterprising environment on the interaction between conscientiousness and neuroticism, in samples with both realistic and enterprising jobs (Jensen & Patel, 2011; King et al., 2005). Studies using mixed samples included a significant share of enterprising jobs (Bowling et al., 2011; Zhou et al., 2014). Extant findings indicate that the enterprising environment activates different outcomes such as helping behaviour (King et al., 2005) and CWB (Bowling et al., 2011; Jensen & Patel, 2011; Zhou et al., 2014). Hence, enterprising environments may also activate the hypothesised interaction between conscientiousness and neuroticism in predicting work passion. Specifically, the interaction between conscientiousness and neuroticism may be stronger for high enterprising than for low enterprising environments, so we hypothesise:

**H4a:** There is a three-way interaction between conscientiousness, neuroticism, and the enterprising environment in predicting HWP. The positive relationship between conscientiousness and HWP will be stronger for low neuroticism and a high enterprising environment than for low neuroticism and a low enterprising environment.

**H4b:** There is a three-way interaction between conscientiousness, neuroticism, and the enterprising environment in predicting OWP. The negative relationship between conscientiousness and OWP will be stronger for low neuroticism and a high enterprising environment than for low neuroticism and a low enterprising environment.

Regarding the influence of enterprising jobs on the interaction between conscientiousness and agreeableness, extant evidence in the literature predominantly found the C*A interaction in samples consisting of enterprising jobs. While most studies found the interaction between conscientiousness and agreeableness to predict performance (Guay et al., 2013, 2013; Warr et al., 2005), there is also evidence that this predictive validity also holds for other outcomes, such as CWB (Jensen & Patel, 2011) or helping behaviour (King et al., 2005). The diversity of outcomes and the fact that enterprising environments activate agreeableness and conscientiousness (Fruyt & Mervielde, 1999; Tett & Burnett, 2003) suggest that the enterprising environment may activate the hypothesised interaction to predict work passion. Hence, the interaction between conscientiousness and agreeableness may be stronger for high enterprising environments than for low enterprising environments. Thus we hypothesise:

**H5a:** There is a three-way interaction between conscientiousness, agreeableness, and enterprising environment in predicting HWP. The positive relationship between conscientiousness and HWP will be stronger for high agreeableness and a high enterprising environment than for high agreeableness and a low enterprising environment.

**H5b:** There is a three-way interaction between conscientiousness, agreeableness, and enterprising environment in predicting OWP. The negative relationship between conscientiousness and OWP will be stronger for high agreeableness and a high enterprising environment than for high agreeableness and a low enterprising environment.

## DATA AND METHOD

### Sample: The myPersonality project

Our sample is the myPersonality project (MPP) data, a large database that mainly entails personality scores and was launched in June 2007 (Kosinski & Stillwell, 2016).

The myPersonality application was available on the social networking platform Facebook and provided psychometric tests and further questionnaires to interested users. In total, some 7.5 million Facebook users completed a MyPersonality questionnaire, whereby the myPersonality database stored six million tests results. Four million users agreed to also share their Facebook profile information (Kosinski & Stillwell, 2016). A low-cost technique for data collection, snowball sampling (Goodman, 1961), recruited participants for the MPP. Snowball sampling accessed the Facebook user base by persuading an initial set of Facebook users to invite their friends to participate who then again invited their friends. When this process reached a critical mass, the sample size rapidly increased (Kosinski et al., 2015). While Facebook facilitates the observation of individuals and can advance social science, it involves novel ethical challenges (Kosinski et al., 2015). Therefore, an extensive consent process preceded the psychometric tests (Kosinski & Stillwell, 2015a).

Kosinski and Stillwell shared the data of the MPP with the academic community on the website http://mypersonality.org up until 2018. Reviewing previous studies working with MPP data implies a gap that concerns the connection of personality data with organisational aspects. This is surprising as previous research found personality to be important for work outcomes, such as performance (e.g. Guay et al., 2013) or CWB (e.g. Zhou et al., 2014).

### Ethics

Throughout the project, we worked with an anonymized dataset that does not contain any information which could lead to the identification of individuals in the sample. The project was reviewed and approved by the University of Oxford's Central University Research Ethics Committee; CUREC no. SSH OII C1A17051.

### The procedure of merging MPP datasets

To explain the impact of personality on passion, we combined MPP personality scores and the questionnaire for work passion. A unique user identifier (ID) allowed tracking individuals across different datasets and enabled the merge of datasets. The personality score dataset was the core of the MPP and encompassed 3.1 million records (Kosinski & Stillwell, 2017b). The passion at work questionnaire available on the MPP website had 1,081 records (Kosinski & Stillwell, 2015b). Merging these two datasets surfaces the participants who answered both the passion and personality questions (*N* = 975).

However, the merged data set lacked information about gender, age, or relationship status. This information was part of the demographics dataset that stored information visible on users' Facebook profiles with 4.3 million records (Kosinski & Stillwell, 2017b). Thus, we conducted a second merge with the demographic dataset. A third merge added the dataset containing employment information (Kosinski & Stillwell, 2017b). The issue of having a large amount of missing data made it difficult to construct a control variable for organisational size and the industry. All control variables were complete for 824 participants.

**Measures**

**Personality.** Personality assessment is the core of the MPP and operationalised with the International Personality Item Pool (IPIP) proxy for the NEO-PI-R, a common measure of the Big Five (Kosinski & Stillwell, 2017a) accessible on the IPIP website. The MPP employed two versions of personality scores: In one version, users completed the 20-100 item IPIP proxy by determining in advance how many items they want to complete or adding questions in blocks of ten items. The chance of feedback on the results leads to reliability greater than .8 (Kosinski & Stillwell, 2017a). 39.4% of participants completed the 100-item version.

A second version was the detailed 336 item IPIP proxy (Costa & McCrae, 1992). Respondents had to complete less attractive experiments to earn credits or paid a little fee (around 4$) to participate (Kosinski & Stillwell, 2017a). This high motivation resulted in reliability measures between .7 to .9 of this scale.[1] 47.8% of the sample completed the 336-item version. To measure the level of a personality trait, participants expressed their level of agreement to items on a 5-point scale (1 = *Strongly disagree*, 5 = *Strongly agree*). An item measuring neuroticism was "Get stressed out easily" (IPIP, 2017b).

**Passion at work scale.** The passion at work questionnaire of the MPP used a six-item work passion scale based on the original passion scale (Vallerand et al., 2003; Vallerand & Houlfort, 2003). The items were measured on a 7-point Likert scale (1 = *Not Agree at All*, 7 = *Very Strongly Agree*). To test the fit of this scale, we performed a confirmatory factor analysis (CFA). The CFA revealed reliability scores resembling previous findings (α of HWP = .89, α of OWP = .80), whereas other fit measures indicated relatively poor fit ($\chi^2$ = 583.141, *df* = 53, *CFI* = .882, *RMSEA* = .115, *SRMR* = .087). To ensure acceptable fit, we, therefore, used a shortened passion scale with four items for each passion type. One of the four items measuring HWP was "My work is in harmony with the other activities in my life", "The new things that I discover with my work allow me to appreciate it even more", "My work reflects the qualities I like about myself", and "My work allows me to live a variety of experiences". An exemplary item for OWP was "I have difficulties controlling my urge to work". Reliability scores for HWP and OWP centred at .82 and .79, respectively.

**RIASEC environments.** The passion questionnaire asked about the job of participants with 24 pre-clustered categories or manual job description options. For classifying the RIASEC environments, this job information was the base. The frame for the classification of RIASEC categories was the Occupational Information Network (O*NET), operated by the US Department of Labor/Employment and Training Administration (O*NET Resource Center, 2017a). The "O*NET interest profiler" characterises each job with its corresponding three-fold RIASEC code (Holland, 1985; O*NET Resource Center, 2017b). As the first letter of the code is most meaningful for determining the vocational match (Holland, 1985), only this letter was used in subsequent analysis. By consulting the interest profiler, we allocated the job from the work passion questionnaire to the RIASEC environment.

**Control variables.** Gender, educational level, and organisational tenure have associations with engagement and/or work performance (Rothbard, 2001; Tsui & O'Reilly, 1989), one of the consequences of passion (e.g. Astakhova, 2015). A binary variable measured gender (1 = female, 2 = male), a scale ranging from 0 (Less than high school) to 6 (Professional Degree) measured the educational level. The scale for organisational tenure ranged from 1 (Less than 6 months) to 7 (21 years or more). We also controlled for the supervisory power of participants (0 = no supervisory power, 1 =

---

[1] It was not possible to determine the exact reliabilities as we only could access the already aggregated personality scores (without access to the underlying items that form this score), Kosinski and Stillwell (2017b).

supervisory power). While supervised employees may be at risk for OWP due to external pressures to comply with directives, the decision autonomy of supervisors may foster HWP (Vallerand & Houlfort, 2003).

We also considered work hours as workers with OWP may work longer than employees with HWP (Vallerand et al., 2003; Vallerand & Houlfort, 2003). Moreover, to capture the superordinate dimension of self-control, we controlled for extraversion and openness to experience as they relate to the superordinate dimension engagement (Olson, 2005). Also, we controlled for the influence of the investigative environment in the regression for the enterprising environment to test for the activating power of the job environment. Both environments show the largest differentiation of job demands (Holland, 1985). The main difference between the investigative and enterprising environments is the absence of social element as well as the emphasis on success and reputation in the former as compared with the latter. In an investigative environment, the focus is on "data" rather than persuasion through social skills. We believe these characteristics make the investigative environment a good choice for the control condition as most of the trait interactions should be originated from ambition and social skills. For each individual in the sample, we coded their job as low/high (0 and 1) on investigative and enterprising.

**Description of the sample**

For 824 participants, all control variables were complete. Regarding organisational tenure, 38% of participants worked less than a year, around a fifth had longer work experience than five years. Even though a lot of information specifically about age was missing (41%), the average age was 27 years ($SD = 10$). The young age of participants may explain the short work experience. One-quarter of respondents had supervisory power. The work hour distribution was bell-shaped and peaked at 31-40h per week. Education level was diverse, nearly a third had some college education ($N = 252$), each fifth participant visited high school/completed a general educational development test (GED) or had a four-year college degree education. 56% of participants were female, 44% male. 340 participants were employed in investigative environments, among them 234 students. 208 participants worked in enterprising jobs. Among the 528 entries about relationship status, 50% were single, 20% were in a relationship and 14% were married. Most of the participants were Caucasian (72%), followed by Asians (11%), Hispanic (6%), and Black (3%). Among 68 countries ($N = 783$), 61% came from the United States, 10% from Great Britain (10%), each 4% from Canada and Australia.

**RESULTS**

**Preliminary analysis**

**Common method variance.** Both the personality and the passion questions were self-reported in the same source, the MPP. This induces the threat of a common method variance problem (Podsakoff, MacKenzie, & Podsakoff, 2012). To test for this bias, we compared the date when respondents completed the passion and personality questions. This relates to the proposed solution of "temporal separation" to address common method variance (Podsakoff, MacKenzie, Lee, & Podsakoff, 2003, p. 887). While a time stamp existed for all passion items ($N = 824$), only 394 time-stamps were available for personality. The dates of these 394 entries were the basis for examining whether there was a temporal separation between the data collection of personality and passion. Among participants, 35 completed the passion and personality questionnaire on the same day, 128 completed first the passion questions followed by the personality scores, and 231 answered the personality before the passion scores. The absolute time difference between the two questionnaires averaged 243 days. Podsakoff et al. (2003) recommended a time difference that prevents the use of previous answers to complete missing information or use short-term memory to answer follow-up questions. The time difference of about eight months seemed sufficiently large as to consider the risk of common method variance small. Similar studies using temporal separation have used similarly long or even shorter intervals in temporal separation (e.g., Ng & Feldman, 2013; Strobel, Tumasjan, Spörrle & Welpe, 2017). The personality data were pre-processed by the MPP so that we only could access the already aggregated trait scores. As the manifest items were only available for passion, we could not test for common method variance bias. This implies the necessity to rely on temporal separation to account for common method variance (Podsakoff et al., 2012).

**Exploratory factor analysis.** First, we conducted an EFA with the passion scale used in the MPP containing 12 items with maximum likelihood as the estimation method. After extracting two factors, we performed a Promax rotation. As a form of oblique rotations, the Promax rotation allows that factors are correlated. Also, HWP and OWP correlated in previous studies (Ho et al., 2011; Vallerand et al., 2003). Results showed a two-factor solution explaining 52.4% of the variable's variance with eigenvalues after the rotation of 3.37 and 2.92 for HWP and OWP, respectively. The items loaded on average with .71 on HWP and .67 on OWP, whereby all loadings were at least .53. One item (q12) showed strong loadings on both factors.

**Confirmatory factor analysis.** To validate the 12-item work passion scale used in the MPP, we performed a CFA with the R package lavaan (Rosseel, 2012) including six passion items for both types of passion. In line with an encompassing simulation study of Beauducel & Wittmann (2005), the comparative fit index (CFI), the standardised root mean square residual (SRMR), and the root means the square error of approximation (RMSEA) served as fit indices. Following propositions in the literature, the cut-off value for CFI was 0.95, for SRMR values close to 0.08 were accepted (Hu & Bentler, 1999, p. 27; Iacobucci, 2010, p. 92). For RMSEA values, 0.06 was the cut-off value (Hu & Bentler, 1999, p. 27). However, the passion scale used in the MPP showed low fit values ($\chi^2 = 583.141$, $df = 53$, $CFI = .882$, $RMSEA = .115$, $SRMR = .087$).

To ensure that the scale has acceptable fit values, we constructed a new passion scale based on the passion scale used in MPP. As the EFA revealed strong secondary loadings of item q12, we excluded this item. Moreover, we kept only items from the MPP scale with the same wording as in the original seven-item passion scale (Vallerand & Houlfort, 2003). Only four MPP items measuring HWP resembled the original, thus we limited the hypothesised number of scale items to four. To explore which item combination yields the best fit for OWP, we combined the one item from the MPP scale that resembled the original wording with other MPP scale items measuring OWP. Table 1 shows the tested models. Our hypothesised new passion at work scale consists of the four MPP items with the original wording for HWP (q1, q3, q5, q6) and the MPP items q4, q2, q9, q11 for OWP. This scale shows acceptable fit ($\chi^2 = 71.464$, $df = 19$, $CFI = .978$, $RMSEA = .060$, $SRMR = .036$). All items had factor loadings greater than .46 on the two passion types and were highly significant ($p = .005$). The internal consistency was similar to the scales reported by Vallerand (2003), with α of OWP = .79 and α of HWP = .82.

-----------------------------------------------
Insert Table 1 about here
-----------------------------------------------

Alternative models 2-4 kept the four items with the original wording for HWP constant and the item q4 for OWP. The hypothesised model fit the data significantly better than model 2 and 4 but was comparable to alternative model 3 ($\chi^2 = 66.241$, $df = 19$, $CFI = .980$, $RMSEA = .056$, $SRMR = .036$). However, an additional EFA revealed that only for the hypothesised model, but not for model 3, a two-factor structure was sufficient to explain the items ($p < .05$). This finding supports the hypothesised model as a two-factor structure is necessary to capture the two types of passion. We additionally tested two five-item scales whereby the items with the original wording remained constant. The first scale (alternative model 5) had better fit than the second but fitted worse compared to the proposed four item-scale ($\chi^2 = 323.245$, $df = 34$, $CFI = .916$, $RMSEA = .105$, $SRMR = .055$). Also, we tested 24 three-item scales but their internal reliability was worse than the four item-subscale (α of HWP between .79 and .82, α of OWP between .74 and .78). These results support the hypothesised four-item scale.

**Descriptive statistics**

Table 2 shows the means and standard deviations of HWP, OWP, the traits, and the control variables as well as their correlations. HWP and OWP were moderately strong, significantly correlated ($r = .56$, $p < .001$). This matches previous findings ($r = .67$, Astakhova, 2015; $r = .46$, Vallerand et al., 2003).

-----------------------------------------------
Insert Table 2 about here
-----------------------------------------------

We tested the hypotheses in three stages. First, we tested the direct relationship between conscientiousness and HWP as well as OWP (Hypothesis 1). Second, we added agreeableness and neuroticism to evaluate how they impact the relation between conscientiousness and passion (Hypotheses 2-3). Third, we integrated the RIASEC environments to account for situational effects (Hypotheses 4-5). We performed hierarchical linear regression on HWP and OWP separately. To minimise multicollinearity, we standardised (z-scored) the three traits to calculate trait interactions (Aiken & West, 1991). In line with the literature, the interaction terms entailed non-standardised binary RIASEC variables (Dawson, 2014). As interaction terms should work with the same variables that enter the regression, also the traits and control variables entered the regression standardised (Dawson, 2014). Table 3 shows the first model that tested the interaction effects for the enterprising environment and Table 4 the second model that controlled for the investigative environment.

-----------------------------------------------
Insert Table 3 about here
-----------------------------------------------
-----------------------------------------------
Insert Table 4 about here
-----------------------------------------------

**Main effects.** Hypothesis 1a and 1b predicted that conscientiousness will positively relate to HWP and negatively to OWP. This hypothesis assumed that the strong self-control capabilities inherent in conscientiousness support HWP but impede OWP. As expected in hypothesis 1a, conscientiousness has a significant, positive relation to HWP ($b = .36$, $p < .001$). Contrary to hypothesis 1b, also the relation between conscientiousness and OWP is significantly positive ($b = .41$, $p < .001$). These results support hypothesis 1a as the positive relation between conscientiousness HWP occurs as predicted. However, the positive relation between conscientiousness and OWP rejects the hypothesised negative relation, thus hypothesis 1b is rejected. The positive relation between conscientiousness and OWP is even stronger than for HWP in both environments.

**Two-way interactions.** Hypotheses 2 and 3 predicted two-way interactions of the hot influence of neuroticism (Hypothesis 2a and 2b) and the cool influence of agreeableness (Hypothesis 3a and 3b) on the conscientiousness and work passion relation.

Hypothesis 2a stated that the positive relation between conscientiousness and HWP will be stronger for low levels of neuroticism than for high levels of neuroticism. Hypothesis 2b predicted the same effect for the negative relation between conscientiousness and OWP. Interactive effects occur for neither HWP ($b = -.06$, $p = .22$) nor OWP ($b = -.02$, $p = .62$). Rejecting both hypotheses reveals that the interaction between conscientiousness and neuroticism is not decisive for predicting work passion across different job environments.

Hypotheses 3a and 3b predicted that conscientiousness will interact with agreeableness, such that the positive relation between conscientiousness an HWP will be stronger for high levels of agreeableness than for low levels of agreeableness (Hypothesis 3a) and the negative relation between conscientiousness and OWP will be stronger for high levels of agreeableness (Hypothesis 3b). As evident from table 3, interactive effects between conscientiousness and agreeableness appear neither for HWP ($b = -.01$, $p = .89$) nor for OWP ($b = -.07$, $p = .13$). These results reject hypotheses 3a and 3b.

A surprising, non-hypothesised interaction occurs between neuroticism and OWP in the enterprising environment ($b = .28$, $p < .05$). In line with the recommendations of Aiken & West (1991), we plotted the significant interaction. High and low levels of the trait were one *SD* above and below the mean, respectively. The binary values of the RIASEC variable determined low and high environments. Figure 2 shows that only for high enterprising environments, high neuroticism increases OWP. A simple slope test (Dawson, 2014) reveals that the coefficient of neuroticism that predicts OWP is positive and significant for the high enterprising environment ($b = .24$, $t = 1.70$ $p = .09$) and slightly negative, but not significant, for the low enterprising environment ($b = -.04$, $t = -.68$, $p = .50$).

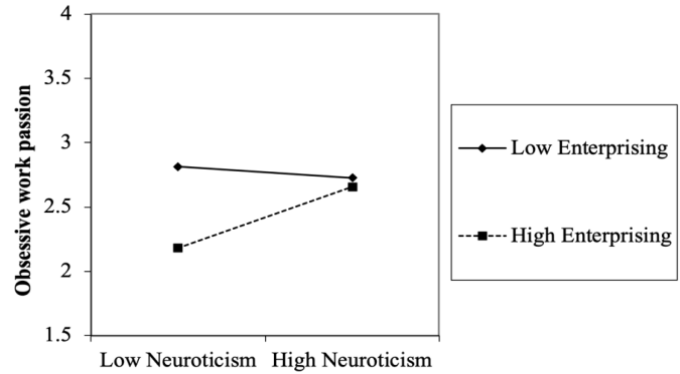

Figure 2: Two-way interaction between neuroticism and the Enterprising environment in predicting OWP with data from model 1.

**Three-way interactions.** Hypotheses 4 and 5 included the enterprising environment. We hypothesized that the enterprising environment will activate the trait interactions such that their effect on predicting HWP and OWP is more pronounced in enterprising jobs.

Hypothesis 4 predicted an interaction between conscientiousness, neuroticism, and the enterprising environment. More precisely, Hypothesis 4a (4b) expected that the positive (negative) relation between conscientiousness and HWP (OWP) is stronger for high enterprising environments and low levels of neuroticism than for low enterprising environments. As shown in Table 3, neither three-way interactions for HWP ($b = -.08$, $p = .49$) nor OWP ($b = .14$, $p = .23$) are significant. This result rejects both hypotheses suggesting that the enterprising environment does not activate these interactions to predict work passion.

However, a significant positive three-way interaction occurs in table 4 between conscientiousness, neuroticism, and the investigative environment in predicting the level of HWP ($b = .22$, $p < .05$). Figure 3 plots this three-way interaction. A simple slope test shows that for the relation between conscientiousness and HWP, the slope of low neuroticism - low investigative (black square) significantly differs from the slope of high neuroticism - low investigative (black diamond), $t = -2.46$, $p = .01$.

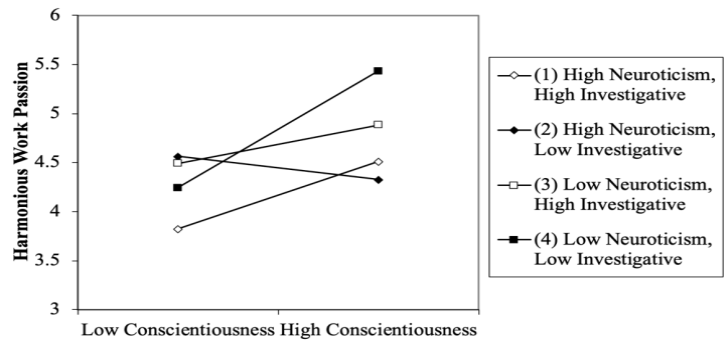

Figure 3: Three-way interaction between conscientiousness, neuroticism, and the Investigative environment predicting HWP, data from model 2.

The slope of low investigative environments is in favour of the hypothesis that neuroticism weakens the relation between conscientiousness and HWP. Paradoxically, high neuroticism strengthens the relation between conscientiousness and HWP in investigative jobs: The slope of high neuroticism - low investigative (black diamond) significantly differs from the slope of high neuroticism - high investigative (white diamond), $t = 1.71$, $p = .09$. This implies that the assumed direction of neuroticism weakening the positive relation between conscientiousness and HWP holds in all work environments but explicitly not in investigative environments.

Hypothesis 5 predicted an interaction between conscientiousness, agreeableness, and the enterprising environment on HWP and OWP. In particular, hypothesis 5a (5b) proposed the positive (negative) relation between conscientiousness and HWP (OWP) to be stronger for high levels of agreeableness and high enterprising than for low enterprising environments. In contrast to hypothesis 5a, no interactive effects occur on predicting HWP ($b = -.11$, $p = .35$). However, table 3 reveals a significant and negative interaction between conscientiousness, agreeableness, and the enterprising environment for OWP ($b = -.22$, $p < .1$). Figure 4 shows that for the relation between conscientiousness and OWP, the slope of low agreeableness - high enterprising (white square) is significantly different from the slope of high agreeableness - high enterprising (white diamond), $t = -2.36$, $p = .02$. As hypothesized, the cooling effect of agreeableness on the relation between conscientiousness and OWP depends on the enterprising environment: The slope of low agreeableness - high enterprising (white square) and low agreeableness - low enterprising (black square) significantly differs ($t = 1.67$, $p = .095$). Low agreeableness significantly strengthens the relation between conscientiousness and OWP, but only in high enterprising environments.

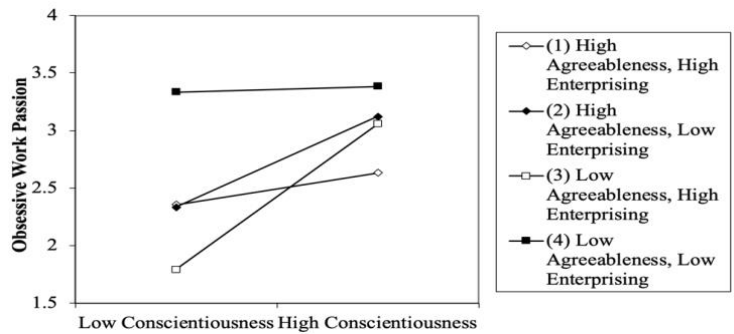

Figure 4: Three-way interaction between conscientiousness, agreeableness, and the Enterprising environment in predicting OWP, data from model 1.

To conclude, the results reject hypothesis 5a but support hypothesis 5b. In particular, for hypothesis 5b, the direction of the interaction is in line with the hypothesised cooling effect of agreeableness on the relation between conscientiousness and OWP.

## DISCUSSION AND CONCLUSION

The goal of this paper was to explore the antecedents of work passion by investigating trait interactions that predict passion at work across different job environments. Integrating passion research with self-regulation theory (Metcalfe & Mischel, 1999) enabled us to make theoretical predictions about the interactions among the three personality traits conscientiousness, agreeableness, and neuroticism. Besides trait interactions, we included situational moderators in form of job environments to enhance the predictive validity of trait interactions (Tett & Burnett, 2003). After a classification of job demands based on Holland's RIASEC taxonomy of vocational choice (1985), we focused on enterprising and investigative job environments because of their trait activating potential perceived in earlier research (e.g. Guay et al., 2013; Tett & Burnett, 2003; Warr et al., 2005).

### Contributions and theoretical implications

**Trait interactions and passion.** First, our paper advances knowledge about the role of personality on work passion by considering the impact of trait interactions. Scholars have long been highlighting the importance of investigating the impact of personality on passion as this realm is still largely unexplored (Perrewé et al., 2014). To the best of our knowledge, previous studies have neglected the impact of personality on work passion. This is surprising as a lot of research recently emerged that studies the consequences of work passion (e.g. Astakhova, 2015; Astakhova & Porter, 2015; Burke et al., 2015). The novel approach of this paper is our focus on the explanatory power and theoretical fruitfulness of trait interactions. Specifically, testing a model with different configurations of the three personality traits conscientiousness, agreeableness, and neuroticism reveals which personality combinations are particularly predictive of HWP and OWP.

Including trait interactions, advances passion research. Previous studies exclusively examined the influence of single personality traits on passion (Balon et al., 2013; Tosun & Lajunen, 2009). A study targeting trait interactions on work involvement, a related construct to work passion (Bozionelos, 2004), found promising avenues for considering trait interactions. Importantly, we find evidence that trait interactions improve the predictive power of traits. Whereas conscientiousness alone positively predicted both HWP and OWP across all contexts, its influence diminished after considering interactive effects.

**Trait activation in job environments.** Second, this paper contributes to understanding how trait interactions affect work passion in different job environments (Tett & Burnett, 2003; Tett & Guterman, 2000) by combining trait *interaction* theory with trait *activation* theory. The enterprising environment was in the focus as it activates all three traits of interest (Fruyt & Mervielde, 1999; Tett & Burnett, 2003). Moreover, by analysing samples of previously conducted trait interactions, we found that trait interactions occurred primarily in enterprising environments (e.g. Guay et al., 2013; Warr et al., 2005). In our analysis regarding work passion, we found mixed support for enterprising jobs activating trait interactions to predict work passion.

The non-significant interaction in the enterprising environment between neuroticism and conscientiousness fits the observation that these trait interactions did not occur in purely enterprising samples (e.g. Jensen & Patel, 2011; King et al., 2005). This result contributes to trait activation theory by suggesting that while a situation might activate a single trait, this activation power does not necessarily hold for trait interactions. However, the enterprising environment activates the interaction between conscientiousness and agreeableness. The situational contingency of the interaction between conscientiousness and agreeableness is in line with the finding that the enterprising environment is particularly prone to keep its activation power also for trait interactions (e.g. Guay et al., 2013; Warr et al., 2005).

Further, the investigative environment activated the interaction between conscientiousness and neuroticism. Investigative environments not only reinforce analytical, intellectual and curious traits but also characteristics such as pessimistic, precise, unpopular, introspective, and cautious (Holland, 1985). These adjectives resemble descriptions of neuroticism (Goldberg, 1990). Trait activation theory formalised these similarities by positing that investigating environments activate neuroticism (Fruyt & Mervielde, 1999; Tett & Burnett, 2003). Thus, both investigative and enterprising environments activate neuroticism (Fruyt & Mervielde, 1999). The results suggest that the investigative environment has a higher activating power with regards to the trait interaction between conscientiousness and neuroticism in predicting work passion. As our results support a systematic influence of job environments on trait interactions, future studies could benefit by considering the job context when investigating the impact of trait interactions.

**Self-regulation theory.** Third, this paper builds on findings by Ode et al. (2008) who used the hot-system/cool-system framework (Metcalfe & Mischel, 1999) to predict the direction of trait interactions between agreeableness and neuroticism. The empirical results supported agreeableness as a cool and neuroticism as a hot influence on the relation between conscientiousness with both HWP and OWP. This extends the findings of Ode et al. (2008) in that the hot and cool influences of neuroticism and agreeableness not only occur while interacting with each other but that they also hold for moderating traits like conscientiousness. Specifically, the positive relation between conscientiousness and HWP was stronger for low than for high neuroticism in low investigative environments. This fits the proposed direction that the hot influencer neuroticism has a decreasing effect on HWP (Vallerand & Houlfort, 2003). The fact that the hypothesised direction occurred for low investigative environments may be due to the characteristics of neuroticism. Neurotic individuals flourish in investigative environments (Holland, 1985; Goldberg, 1990), thus neuroticism may enhance HWP.

When activated by the enterprising environment, agreeableness reduces the positive relation between conscientiousness and OWP such that the relationship was stronger for low agreeableness than for high agreeableness. This is in line with the proposed direction that agreeableness as a cool influencer decreases OWP. The high self-control abilities of agreeableness contrast OWP which relates to a control loss of work (Vallerand & Houlfort, 2003).

Our findings also enhance research dealing with the relation between personality traits and self-control. Extant results support the impulsivity of neuroticism, the self-control abilities of agreeableness, and the ambivalence of conscientiousness being related to both high self-control and deliberation (McCrae & Löckenhoff, 2010; Whiteside & Lynam, 2001). Contrary to the hypotheses that the deliberate forces of conscientiousness prevent OWP, it positively predicted both HWP and OWP. The deliberation facet may be too weak to stop the rigid self-discipline in favour of OWP. This ambiguity of conscientiousness as a double-edged sword fits previous findings (McCrae & Löckenhoff, 2010).

**Practical implications**

The relatively low number of passionate employees in organisations fuels the need to understand what makes employees passionate about their work. Although we found evidence that the interactions occurred only in specific job environments, the increase in explained variance of the model was small. This implies that personality contributes a small, but significant part to explain passion at work and depends on the work environment.

The first practical implication relates to the selection and recruiting efforts. While enterprising and investigative jobs reported different levels of HWP, all three traits were relatively constant independent of the job environment. Due to the relatively stable disposition of personality, organisations may include personality measures as a selection criterion to explore its impact on passion. The ethical consequences of this approach would require an intensive discussion about further usage, storage, and the accessibility of this personality data. Before restructuring established selection processes, further research should validate how personality predicts work passion. It may result in an extensive catalogue similar to the O*NET that entails information about personality combinations predicting HWP and/or OWP for all jobs.

A second implication refers to coaching employees with an increased risk of OWP. This enables fast implementation and helps organisations to maximise the potential of the workforce employed at present. Personality-tailored coaching could specifically target employees with an increased risk for OWP, such as low agreeable, high conscientious employees in enterprising jobs. The high propensity of OWP to burnout (Vallerand et al., 2010) could be the first area of coaching. It may focus on habits of employees with HWP that prevent burnout, such as keeping work and leisure apart or not work overtime (Vallerand et al., 2010).

Third, organisations may consider the situational strength of job environments. As this paper showed, low agreeable, high conscientious employees are at increased risk for OWP in enterprising jobs. In strongly directed jobs, the risk for OWP might be low as employees have little power to deviate from directions. In jobs with much autonomy, individual differences become apparent and OWP is more likely. Organisations may improve workplace environments that foster tendencies of employees for HWP but lowers them for OWP.

**Limitations and future research ideas**

This paper has several limitations. The first limitation concerns the data as working with data from online social networks is still relatively rare in social sciences (Kosinski et al, 2015). Primarily, the MPP used the snowball sampling method to collect the data that induced a selection bias: Besides individuals tend to interact with similar people, well-connected people are easier recruited (Illenberger & Flötteröd, 2012). Nevertheless, a comparison of the personality scores collected with the MPP and traditional pen and paper experiments revealed a similar selection bias (Kosinski et al., 2015) which extenuates the selection bias argument. A second constraint of the data was that answers were self-reported from questionnaires available on the MPP. However, the results of a common method variance analysis revealed that common method variance was not a problem for this paper: The average difference between completing personality and passion items seemed long enough to minimise common method variance (Podsakoff et al., 2012). A last limitation of the data concerned the passion at work scale from the MPP. Due to the bad fit values of the MPP scale, we developed a shortened version of this passion at the work scale. This new scale would require further validation with an independent sample or a validation questionnaire, however, it seemed more appropriate to construct a new scale than to build our analysis on an unreliable scale.

Second, this paper applied self-classified RIASEC environments to account for the activating effect of situations. As no previous research clustered trait-interaction samples into RIASEC types, we decided to cluster the jobs indicated by participants in RIASEC environments by consulting the "O*NET interest profiler". Run by the US Department of Labor/Employment and Training Administration (O*NET Resource Center, 2017a), the O*NET serves as an official facilitator of characterising jobs to RIASEC environments. Future studies should replicate our clustering to achieve further classification validity. This classification may fuel research testing the influence of personality traits on passion across other jobs than the enterprising and investigative environment. Notably, investigating one specific trait interaction across all six RIASEC environments might provide valuable insights. It may contribute to the vision of creating a dictionary informing about which trait combinations predict HWP or OWP for each job. Large scale online data such as the MPP has the potential to accelerate these efforts as a cheap, reliable source of heterogeneous samples for social science research.

Third, we only investigated interactions between conscientiousness, agreeableness, and neuroticism on HWP and OWP. Acknowledging that other interactions also might yield interesting insights, we chose these traits due to their affiliation to the superordinate trait dimension of self-control (Olson, 2005). Research could benefit by considering not only trait pairs in interaction with a context but also trait triples. Also, the traits of the superordinate dimension engagement extraversion and openness may significantly impact passion. Further studies may compare the differences between the two superordinate dimensions and their predictive power regarding work passion. When comparing the interaction effects of different trait combinations on passion, it would be helpful to test these for one specific context first. Focusing first on the enterprising or investigative environment could be a fruitful avenue to deepen the understanding of how trait interactions impact work passion.

Our study advances the understanding of the antecedents of work passion by focusing on trait interactions. In addition, this paper includes job environments to compare the effect of trait interactions on work passion across different jobs. We found that the enterprising environment activates the interaction between conscientiousness and agreeableness on OWP. Moreover, the investigative environment activates the interaction between conscientiousness and neuroticism on HWP. Thereby, the direction of the observed interactions fits the notion that agreeableness is a cool influencer on the relation between conscientiousness and OWP, and neuroticism is a hot influencer on the relation between conscientiousness and HWP. We hope that this paper fuels further, fruitful research based on large scale online data that deepens the knowledge on how traits interact in different job environments to predict passion at work.

## ACKNOWLEDGMENT
We thank Michal Kosinski and David Stillwell for sharing the myPersonality data with us. We thank Otto Kässi for his useful discussions, and Andranik Tumasjan for the help with the analysis. We thank Isabell Welpe for her advice on the project.

## FUNDING
TY was partially supported by the Alan Turing Institute under the EPSRC grant no. EP/N510129/1. The sponsor had no role in study design; in the collection, analysis and interpretation of data; in the writing of the report; and in the decision to submit the article for publication.

## CONFLICT OF INTEREST
The authors declare that they have no conflict of interest.

## ETHICS APPROVALL
The project was reviewed and approved by the University of Oxford's Central University Research Ethics Committee; CUREC no. SSH OII C1A17051.

## DATA AVAILABILITY
The datasets analysed during the current study are available from the corresponding author on reasonable request.

Table 1: Alternative models in comparison to the hypothesised model from CFA

| Measurement model | $\chi^2$ | df | $\chi^2/df$ | $\Delta\chi^2$ | $\Delta df$ | CFI | RMSEA | SRMR |
|---|---|---|---|---|---|---|---|---|
| **Hypothesised model:** four-item model (1,3,5,6 – 2,4,9,11) loading on two factors | 71.464 | 19 | 3.761 | | | .978 | .060 | .036 |
| **Alternative model 1**: six-item model (1,3,5,6,8,10 – 2,4,7,9, 11,12) loading on two factors | 583.141 | 53 | 11.002 | 511.680* | 34 | .882 | .114 | .087 |
| **Alternative model 2:** four-item model (1,3,5,6 – 2,4,7,9) loading on two factors | 150.961 | 19 | 7.945 | 79.496* | 0 | .946 | .095 | .042 |
| **Alternative model 3:** four-item model (1,3,5,6 – 2,4,9,11) loading on two factors | 66.241 | 19 | 3.486 | -5.223 | 0 | .98 | .056 | .036 |
| **Alternative model 4:** four-item model (1,3,5,6 – 4,7,9,11) loading on two factors | 115.674 | 19 | 6.088 | 44.21* | 0 | .962 | .081 | .040 |
| **Alternative model 5:** best five-item model (1,3,5,6,8 – 2,4,7, 9,11) loading on two factors | 323.245 | 34 | 9.507 | 251.780* | 15 | .916 | .105 | .055 |

*Note*. * p < .001. Alternative models in comparison to the hypothesised model. The original item number from the items of the work passion scale from the MPP is in the bracket

Table: 2 Means, standard deviations, and correlations

| Variable | M | SD | 1 | 2 | 3 | 4 | 5 | 6 | 7 | 8 | 9 | 10 | 11 | 12 | 13 |
|---|---|---|---|---|---|---|---|---|---|---|---|---|---|---|---|
| 1. HWP | 4.51 | 1.53 | .82 | | | | | | | | | | | | |
| 2. OWP | 2.68 | 1.42 | .56*** | .79 | | | | | | | | | | | |
| 3. Con | 3.51 | .73 | .27*** | .23*** | - | | | | | | | | | | |
| 4. Agr | 3.55 | .61 | .16*** | -.04 | .12*** | - | | | | | | | | | |
| 5. Neu | 2.73 | .82 | -.27*** | -.05 | -.33*** | -.34*** | - | | | | | | | | |
| 6. Gender | 1.44 | .50 | -.01 | .03 | -.01 | -.07* | -.20** | - | | | | | | | |
| 7. Education | 2.44 | 1.53 | .11** | .06* | .15*** | -.02 | -.07* | .03 | - | | | | | | |
| 8. Position | .24 | .43 | .17*** | .14** | .17*** | .04 | -.10** | -.02 | .13*** | - | | | | | |
| 9. Workh | 2.76 | 1.22 | .09** | .14*** | .22*** | .05 | -.09* | .03 | .21*** | .19*** | - | | | | |
| 10. OT | 3.20 | 1.63 | .00 | .01 | -.08* | .08** | -.07† | -.07† | -.05 | .14*** | .27*** | - | | | |
| 11. E | .25 | .43 | -.20*** | -.07* | .02 | -.00 | -.04 | .00 | .08* | .13*** | .09* | .02 | - | | |
| 12. I | .41 | .50 | .00 | -.01 | -.05 | -.06† | .06† | .05 | -.16*** | -.18*** | -.10** | .10** | -.49*** | - | |
| 13. Ope | 4.12 | .55 | .17*** | .09* | .04 | .11** | -.10** | -.04 | .10** | .04 | .03 | -.02 | -.03 | .02 | - |
| 14. Ext | 3.13 | .83 | .21*** | .04 | .19*** | .28*** | -.40*** | -.01 | .06† | .17*** | .10** | .01 | .08* | -.13*** | .22*** |

*Note:* $N = 824$. † $p < .1$. * $p < .05$. ** $p < .01$. *** $p < .001$. HWP = Harmonious Work Passion; OWP = Obsessive Work Passion; Con = Conscientiousness; Agr = Agreeableness; Neu = Neuroticism; Workh = Workhours; OT = Organisational Tenure; E = Enterprising environment; I = Investigative environment; Ope = Openness to experience; Ext = Extraversion. Cronbach's α of the two passion types are on the diagonal.

Table 3: Regression result for the effect of agreeableness, neuroticism, and enterprising environments on the conscientiousness - passion relation

| Variable | Model 1. Enterprising environment | | | | | | | |
|---|---|---|---|---|---|---|---|---|
| | Harmonious Passion | | | | Obsessive Passion | | | |
| | Step 1 | Step 2 | Step 3 | Step 4 | Step 1 | Step 2 | Step 3 | Step 4 |
| Controls | | | | | | | | |
| Gender | -.02 | -.05 | -.05 | -.05 | .04 | .04 | .04 | .05 |
| Education | .11* | .07 | .08 | .07 | .01 | -.02 | -.03 | -.03 |
| Position type | .20*** | .19*** | .19*** | .19*** | .18*** | .15** | .16** | .16** |
| Work hours | .08 | .04 | .04 | .04 | .18*** | .14** | .14** | .14** |
| Organisational tenure | -.05 | -.07 | -.08 | -.08 | -.05 | -.05 | -.06 | -.07 |
| Openness | .19*** | .17*** | .17*** | .17*** | .12* | .13* | .12* | .12* |
| Extraversion | .24*** | .09† | .09† | .10† | -.01 | -.02 | -.02 | -.02 |
| Investigative | .09† | -.10† | -.09 | -.09 | .03 | -.05 | -.05 | -.05 |
| Step 2: Main effects | | | | | | | | |
| Conscientiousness | | .26*** | .24*** | .24*** | | .30*** | .29*** | .30*** |
| Agreeableness | | .07 | .08 | .08 | | -.09† | -.14* | -.14* |
| Neuroticism | | -.24*** | -.29*** | -.28*** | | -.03 | -.04 | -.04 |
| Enterprising | | -.92*** | -.90*** | -.92*** | | -.35** | -.35** | -.29* |
| Step 3: Two-way interaction | | | | | | | | |
| C*A | | | -.01 | -.02 | | | -.07 | -.03 |
| C*N | | | -.06 | -.04 | | | -.02 | -.05 |
| E*C | | | .07 | .09 | | | .05 | .09 |
| E*A | | | -.03 | -.03 | | | .17 | .18 |
| E*N | | | .16 | .17 | | | .28* | .32* |
| Step 4: Three-way interaction | | | | | | | | |
| E*C*A | | | | -.11 | | | | -.22† |
| E*C*N | | | | -.08 | | | | .14 |
| R² | .09 | .21 | .21 | .21 | .04 | .09 | .10 | .11 |
| Adjusted R² | .08 | .20 | .20 | .19 | .03 | .08 | .08 | .09 |
| ΔR² | .09*** | .12*** | .00 | .00 | .04*** | .05*** | .01 | .01* |

*Note.* N = 824. † $p < .1$. * $p < .05$. ** $p < .01$. *** $p < .001$. C = Conscientiousness; A = Agreeableness; N = Neuroticism; E = Enterprising.

Table 4: Regression result for the effect of agreeableness, neuroticism, and investigative environment on the conscientiousness - passion relation

| Variable | Model 2. Investigative environment | | | | | | | |
|---|---|---|---|---|---|---|---|---|
| | Harmonious Passion | | | | Obsessive Passion | | | |
| | Step 1 | Step 2 | Step 3 | Step 4 | Step 1 | Step 2 | Step 3 | Step 4 |
| Controls | | | | | | | | |
| Gender | -.01 | -.05 | -.05 | -.06 | .04 | .04 | .05 | .05 |
| Education | .12* | .07 | .07 | .07 | .01 | -.02 | -.03 | -.03 |
| Position type | .23*** | .19*** | .18*** | .18** | .19*** | .15** | .15** | .16** |
| Work hours | .09† | .04 | .04 | .05 | .19*** | .14** | .14** | .14** |
| Organisational tenure | -.06 | -.07 | -.08 | -.07 | -.06 | -.05 | -.05 | -.06 |
| Openness | .17*** | .17*** | .17*** | .18*** | .11* | .13* | .12* | .12* |
| Extraversion | .26*** | .09† | .09† | .10† | -.01 | -.02 | -.02 | -.02 |
| Enterprising setting | -.36*** | -.40*** | -.39*** | -.39*** | -.14** | -.15** | -.16** | -.16** |
| Step 2: Main effects | | | | | | | | |
| Conscientiousness | | .26*** | .25*** | .26*** | | .30*** | .30*** | .30*** |
| Agreeableness | | .07 | .08 | .07 | | -.09† | -.06 | -.05 |
| Neuroticism | | -.24*** | -.22** | -.23** | | .03 | .09 | .09 |
| Investigative | | -.20† | -.18 | -.11*** | | -.10 | -.10 | -.12 |
| Step 3: Two-way interaction | | | | | | | | |
| C*A | | | -.00 | .01 | | | -.07 | -.08 |
| C*N | | | -.06 | -.14* | | | -.02 | -.00 |
| I*C | | | .03 | .02 | | | .02 | .02 |
| I*A | | | -.03 | -.01 | | | -.09 | -.09 |
| I*N | | | -.06 | -.03 | | | .13 | .14 |
| Step 4: Three-way interaction | | | | | | | | |
| I*C*A | | | | -.04 | | | | .04 |
| I*C*N | | | | .22* | | | | -.05 |
| R² | .14 | .21 | .21 | .22 | .05 | .09 | .10 | .10 |
| Adjusted R² | .13 | .20 | .19 | .20 | .04 | .08 | .08 | .08 |
| ΔR² | .14*** | .07*** | .00 | .01† | .05*** | .04*** | .01 | .00 |

*Note.* N = 824. † p < .1. * p < .05. ** p < .01. *** p < .001. C = Conscientiousness; A = Agreeableness; N = Neuroticism; I = Investigative.